\documentclass[10pt,twocolumn,letterpaper]{article}

\usepackage{cvpr}              

\usepackage{amsmath,amssymb,amsthm}
\usepackage{algorithm}
\usepackage{algorithmic}
\usepackage{booktabs}
\usepackage{graphicx}
\usepackage{tabularx}
\usepackage{multirow}
\usepackage{subcaption}
\usepackage{xcolor}
\usepackage{url}
\usepackage{fontawesome}

\usepackage{booktabs}
\usepackage{siunitx}
\definecolor{cvprblue}{rgb}{0.21,0.49,0.74}
\usepackage[pagebackref,breaklinks,colorlinks,allcolors=cvprblue]{hyperref}

\def\paperID{32} 
\def\confName{CVPR}
\def\confYear{2026}

\title{\textbf{Neuro-Oracle: A Trajectory-Aware Agentic RAG Framework for Interpretable Epilepsy Surgical Prognosis}}

\author{
Aizierjiang Aiersilan \textsuperscript{\faEnvelopeO} \quad
Mohamad Koubeissi\\
The George Washington University \\
{\tt\small alexandera@gwu.edu}
}

\begin{document}
 \maketitle

\begin{abstract}
Predicting post-surgical seizure outcomes in pharmacoresistant epilepsy is a clinical challenge. Conventional deep-learning approaches operate on static, single-timepoint pre-operative scans, omitting longitudinal morphological changes. We propose \emph{Neuro-Oracle}, a three-stage framework that: (i)~distils pre-to-post-operative MRI changes into a compact 512-dimensional trajectory vector using a 3D Siamese contrastive encoder; (ii)~retrieves historically similar surgical trajectories from a population archive via nearest-neighbour search; and (iii)~synthesises a natural-language prognosis grounded in the retrieved evidence using a quantized Llama-3-8B reasoning agent.
Evaluations are conducted on the public EPISURG dataset ($N{=}268$ longitudinally paired cases) using five-fold stratified cross-validation. Since ground-truth seizure-freedom scores are unavailable, we utilize a clinical proxy label based on the resection type. We acknowledge that the network representations may potentially learn the anatomical features of the resection cavities (i.e., temporal versus non-temporal locations) rather than true prognostic morphometry. Our current evaluation thus serves mainly as a proof-of-concept for the trajectory-aware retrieval architecture.
Trajectory-based classifiers achieve AUC values between 0.834 and 0.905, compared with 0.793 for a single-timepoint ResNet-50 baseline. The Neuro-Oracle agent (M5) matches the AUC of purely discriminative trajectory classifiers (0.867) while producing structured justifications with zero observed hallucinations under our audit protocol. A Siamese Diversity Ensemble (M6) of trajectory-space classifiers attains an AUC of 0.905 without language-model overhead.
\end{abstract}

\section{Introduction}
\label{sec:intro}

Epilepsy is a chronic neurological disorder characterised by recurrent, unprovoked seizures, affecting an estimated 50~million individuals globally~\cite{wu2025review}.
Anti-seizure medications achieve adequate control in approximately 70\% of patients; the remaining 30\% develop pharmacoresistance, for whom resective neurosurgery represents the principal curative treatment route~\cite{wiebe2001surgery}.
Temporal lobectomy yields seizure freedom in roughly 65--70\% of operated patients at two years (Engel Class~I), whereas non-temporal and multilobar resections achieve freedom in only 40--55\% of cases~\cite{spencer2008outcomes,engel2003practice,tellez2005long}.
Long-term follow-up data demonstrate that even among initial responders, a substantial fraction experience late relapse~\cite{detisi2011longterm}.
This gap in outcomes across surgery types and the challenge of identifying patients most likely to benefit motivate the search for reliable pre-operative prognostic tools.

Accurate prognosis is especially difficult in patients with structurally normal MRI scans, where the epileptogenic tissue produces no gross visible lesion~\cite{sanders2024outcome}.
In these cases, pathology may manifest as subtle, progressive cortical thinning that evolves across longitudinal imaging acquisitions~\cite{bernhardt2009longitudinal}.
Three structural challenges compound the difficulty of applying deep learning to this setting.
First, static single-scan classifiers evaluate the anatomy at a single cross-section in time and therefore cannot represent the trajectory of anatomical change from a healthy baseline through to the post-operative state.
Second, epilepsy surgical cohorts at individual centres are inherently small, rarely exceeding a few hundred patients with complete imaging and outcome records, which makes high-capacity end-to-end classifiers susceptible to overfitting~\cite{litjens2017survey}.
Third, the output of a standard classifier is a bare probability score carrying no verifiable clinical rationale, which limits its acceptability in surgical decision support.

We address these three challenges jointly with \emph{Neuro-Oracle}.
The system encodes the morphological delta between paired pre-operative and post-operative T1-weighted MRIs into a compact trajectory vector via a 3D Siamese contrastive network~\cite{chopra2005learning,khosla2020supcon}, retrieves the most geometrically similar historical trajectories from a population archive through exact cosine nearest-neighbour search~\cite{johnson2019billion}, and applies a quantized Llama-3-8B reasoning agent~\cite{grattafiori2024llama} to produce a verdict grounded in the retrieved evidence, following the retrieval-augmented generation (RAG) paradigm~\cite{lewis2020retrieval}.
Every component relies exclusively on public-domain datasets and open-source tooling, ensuring reproducibility.

We use EPISURG~\cite{episurg2020}, a public collection of T1-weighted MRIs from epilepsy surgery patients, for all our experiments in this paper (Section~\ref{sec:data}).

The remainder of the paper reviews related work (Section~\ref{sec:related}), describes the dataset and preprocessing (Section~\ref{sec:data}), develops the methodology (Section~\ref{sec:method}), and presents results (Section~\ref{sec:results}) followed by discussion and limitations (Section~\ref{sec:discussion}).

\section{Related Work}
\label{sec:related}

\paragraph{Deep learning for epilepsy prognosis.}
Early computational methods for surgical outcome prediction relied on hand-crafted morphometric features, cortical thickness, hippocampal volume, and gyrification indices, fed into support-vector machines or random forests~\cite{wu2025review}.

3D CNN architectures, notably residual networks~\cite{he2016resnet}, have subsequently been applied directly to raw T1-weighted volumes to learn prognosis representations, demonstrating reasonable internal accuracy on single-centre cohorts~\cite{litjens2017survey}.
Graph neural networks have also been explored to capture the spatial topology of white-matter connectivity in relation to seizure outcomes.
A recurring limitation across these approaches is that they operate on a single pre-operative scan, discarding the clinically informative signal encoded in longitudinal change.
Moreover, class imbalance and cohort sizes of only a few hundred patients render end-to-end classifiers vulnerable to overfitting.

\paragraph{Longitudinal and deformation-based neuroimaging.}
Longitudinal structural MRI analysis has a long history in neurodegeneration research, where within-subject template estimation enables unbiased measurement of volumetric change~\cite{reuter2012within,bernhardt2009longitudinal}.
Siamese and twin-branch architectures, originally introduced for signature verification and metric learning~\cite{chopra2005learning}, have been adopted for longitudinal change detection in conditions such as multiple-sclerosis lesion evolution and tumour volume tracking.
Deformation-based morphometry characterises anatomical change through non-linear registration fields produced by tools such as ANTs~\cite{avants2011reproducible}; however, deformable registration risks mathematically smoothing the localised volume changes of interest in epilepsy.
Neuro-Oracle instead utilizes rigid coregistration followed by a learned difference embedding, preserving the structural signal while representing it in a low-dimensional, retrieval-compatible space.

\paragraph{3D medical-image transformers.}
The Vision Transformer~\cite{dosovitskiy2020image} has rapidly been adapted to volumetric medical data.
UNETR~\cite{hatamizadeh2022unetr} demonstrated that a transformer encoder can serve as the contracting path of a 3D segmentation architecture, and MONAI~\cite{cardoso2022monai} provides reference implementations for both 3D ResNets and 3D ViTs in the medical imaging domain.
In our experimental setup, a MONAI 3D ViT serves as a strong static baseline (M2), allowing direct comparison between transformers operating on single-timepoint inputs and our trajectory-based approach.

\paragraph{Contrastive representation learning.}
Self-supervised contrastive objectives such as SimCLR~\cite{chen2020simclr} and MoCo~\cite{he2020moco} learn discriminative representations by maximising agreement between augmented views of the same instance.

Khosla~et~al.~\cite{khosla2020supcon} extended the InfoNCE loss to the supervised setting, leveraging label information to form positive pairs across samples of the same class.
We adopt Supervised Contrastive loss at temperature $\tau{=}0.07$ and combine it with Focal Loss~\cite{lin2017focal} to jointly address class separation and the 4:1 label imbalance in EPISURG.

\paragraph{Retrieval-augmented generation in clinical settings.}
RAG~\cite{lewis2020retrieval} grounds LLM outputs in an external knowledge store retrieved at inference time, mitigating hallucination and providing traceable evidence chains.
Vector search frameworks like FAISS~\cite{johnson2019billion} allow these archives to scale seamlessly to large datasets.
Extending the RAG paradigm to vision--language clinical applications is an active frontier: MMed-RAG~\cite{xia2024mmed} introduced multimodal retrieval for medical VLMs, and RegioMix~\cite{yung2024region} applied region-specific retrieval augmentation for longitudinal visual question answering.
To the best of our knowledge, agentic visual-RAG, in which the retrieved evidence is processed by an LLM agent rather than a fixed classifier, has not previously been applied to epilepsy surgical prognosis with 3D volumetric longitudinal trajectories as the retrieval key.

\paragraph{Large language models for medicine.}
Foundation language models~\cite{vaswani2017attention,brown2020language} have achieved strong performance on medical question answering, with Med-PaLM~\cite{singhal2023large} reaching expert-level accuracy on the USMLE benchmark.
The open-weight LLaMA family~\cite{touvron2023llama,grattafiori2024llama} and efficient quantisation techniques such as NF4~\cite{dettmers2022qlora} have made it feasible to deploy 8B-parameter models on a single consumer GPU (${\sim}6$\,GB VRAM).
We leverage Llama-3-8B-Instruct in 4-bit NF4 mode as the reasoning agent, loaded via the HuggingFace Transformers library~\cite{wolf2020transformers}.

\section{Datasets and Preprocessing Protocol}
\label{sec:data}

\paragraph{EPISURG.}
The primary dataset we use in this work is EPISURG~\cite{episurg2020}, which contains 430 T1-weighted MRI scans from epilepsy surgery patients, out of which 268 possess both a pre-operative scan $X_{t_0}$ and a post-operative scan $X_{t_1}$.
The public data release provides paired NIfTI volumes alongside a metadata file recording each subject's surgery type, hemisphere, and a pre-operative quality flag; readers can find further dataset details in the original publication~\cite{episurg2020}.
Per-patient Engel seizure-freedom scores are not included in the public release.

Because ground-truth seizure-freedom outcomes are unavailable, we assign a binary proxy label derived from surgery type according to the published literature on outcome rates by surgery category~\cite{wiebe2001surgery,spencer2008outcomes,engel2003practice,tellez2005long}.
Temporal resections (including lobectomy, lesionectomy, and multilobar temporal resections) are labelled $y{=}0$ (favourable outcome proxy, cited seizure-freedom rate 65--70\% at two years), and purely extra-temporal, frontal, parietal, occipital resections, and hemispherectomy are labelled $y{=}1$ (less-favourable outcome proxy, cited rate 40--55\%).
This proxy is a clinically grounded but imperfect approximation. Because the model distinguishes temporal from non-temporal procedures using post-operative scans, the encoder may primarily learn the anatomical features of temporal lobectomy cavities rather than subtle epileptogenic pathology. Consequently, the proxy introduces a major vulnerability where the model predicts surgery type instead of true seizure outcome. All results must therefore be strictly understood as performance on this proxy classification task. The proposed Neuro-Oracle framework serves primarily as an architectural proof-of-concept, designed to be readily applicable once true per-patient outcome registries become available.

After preprocessing quality filtering, \textbf{268 subjects} are retained: 215 with proxy label $y{=}0$ and 53 with $y{=}1$, yielding a case-to-control ratio of approximately 4:1.
While on-the-fly random augmentations are applied during training to prevent overfitting (detailed in Section~\ref{sec:encoding}), no offline augmented or synthetic subject records are added to the overall dataset count.


\paragraph{Preprocessing pipeline.}
Accurate trajectory encoding depends on geometric alignment between timepoints.
Each pre-operative volume $X_{t_0}$ and its corresponding post-operative volume $X_{t_1}$ are read via ANTsPy~\cite{avants2011reproducible}.
The post-operative scan is rigidly registered (6 degrees of freedom: three translations and three rotations) into the physical space of the pre-operative scan:
\begin{equation}
X_{t_1}^{\text{reg}} = \mathcal{T}_{\text{rigid}}\!\left(X_{t_1};\;X_{t_0}\right),
\end{equation}
where $\mathcal{T}_{\text{rigid}}$ is the optimal rigid (6-DOF) alignment mapping $X_{t_1}$ into the coordinate space of $X_{t_0}$.
Deformable registration is deliberately avoided, as non-linear warping would smooth the localised volume changes targeted by prediction~\cite{reuter2012within}.
Both volumes are then z-score normalised per-volume over brain voxels exceeding the per-volume mean intensity (a simple threshold mask sufficient for well-normalised T1w acquisitions), and are centre-cropped or symmetrically zero-padded to a uniform $128 \times 128 \times 128$ voxel grid at $1 \times 1 \times 1$\,mm$^3$ isotropic resolution.
Processed volumes are saved as NIfTI files with an identity affine.
All 268 subjects passed quality control and were carried forward to the experimental pipeline.

\section{Methodology}
\label{sec:method}

\begin{figure*}[t]
    \centering
    \includegraphics[width=\linewidth]{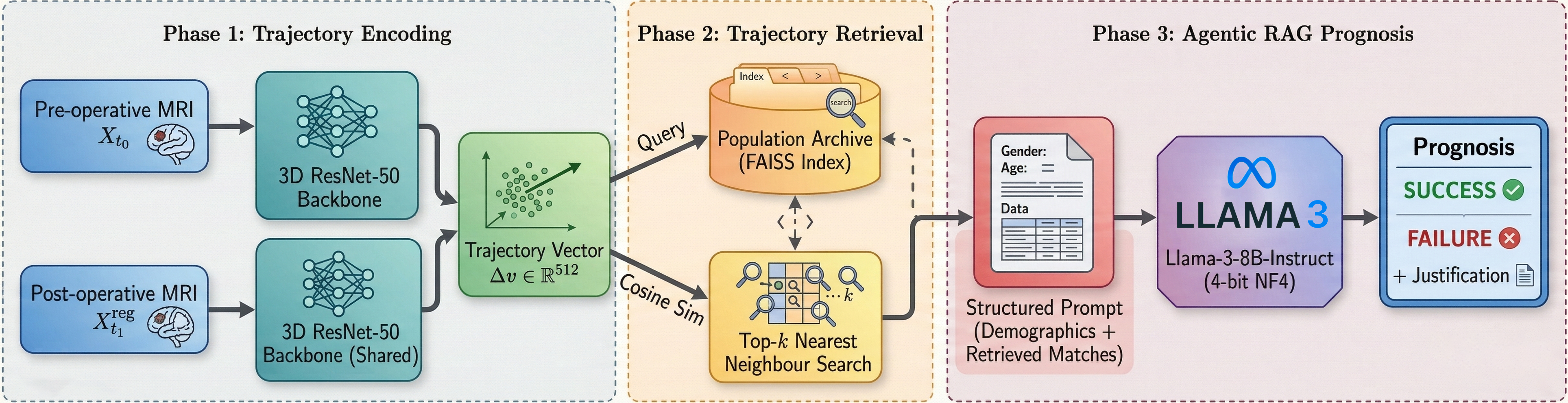}
    \caption{Overview of the proposed Neuro-Oracle framework. (i) Pre- and post-operative MRI pairs are registered and encoded through a shared 3D Siamese ResNet-50 backbone. (ii) The morphological difference is projected into a compact trajectory vector, which is used to retrieve historically similar cases from a population archive. (iii) A quantized LLM reasoning agent evaluates the retrieved evidence to synthesize a transparent, natural-language surgical prognosis.}
    \label{fig:framework}
\end{figure*}

We formulate the prognosis task as follows: given the aligned scan pair $(X_{t_0}, X_{t_1}^{\text{reg}})$ and the surgery-type proxy label $y \in \{0,1\}$, learn a representation of the longitudinal change that separates the two outcome classes.
The pipeline proceeds in three phases that can be trained and deployed independently (see Figure~\ref{fig:framework}).

\subsection{Phase 1: Contrastive Trajectory Encoding}
\label{sec:encoding}

A \textbf{3D Siamese contrastive encoder} shares weights between the two temporal branches~\cite{chopra2005learning}.
The backbone $f_\theta$ is a MONAI~\cite{cardoso2022monai} 3D ResNet-50~\cite{he2016resnet} (bottleneck blocks, layers $[3,4,6,3]$, channel widths $[64,128,256,512]$, single-channel input) initialised from scratch.
All experiments were executed on computer of an NVIDIA GeForce RTX 5060 GPU.

Each scan pair is encoded independently by the shared backbone:
\begin{equation}
v_{t_0} = f_\theta(X_{t_0}), \qquad v_{t_1} = f_\theta(X_{t_1}^{\text{reg}}).
\end{equation}

The difference vector $v_{t_1} - v_{t_0}$ is projected to a 512-dimensional space by an MLP head and $\ell_2$-normalised to produce the \emph{trajectory vector}:
\begin{equation}
\Delta v = \ell_2\text{-norm}\!\left(\text{MLP}(v_{t_1} - v_{t_0})\right) \in \mathbb{R}^{512},
\end{equation}
where $\text{MLP} = \text{Linear}(2048{\to}512) \to \text{BN} \to \text{ReLU} \to \text{Dropout}(0.3) \to \text{Linear}(512{\to}512)$.

\paragraph{Training objective.}
The encoder is jointly trained with two complementary loss terms in equal weight:
\begin{equation}
\mathcal{L} = \tfrac{1}{2}\,\mathcal{L}_{\text{SupCon}} + \tfrac{1}{2}\,\mathcal{L}_{\text{Focal}}.
\end{equation}

The Supervised Contrastive loss~\cite{khosla2020supcon} encourages trajectory vectors from the same surgical-outcome class to have high inner product and those from different classes to be mutually repulsive.
Let $I$ denote the in-batch indices, $P(i)$ the set of positive examples for sample $i$ (same class, excluding $i$ itself), and $\tau{=}0.07$ the temperature:
\begin{equation}
\begin{split}
\mathcal{L}_{\text{SupCon}} &= \sum_{i \in I} \frac{-1}{|P(i)|} \sum_{p \in P(i)} \\
&\quad \log \frac{ \exp(\Delta v_i \cdot \Delta v_p \,/\, \tau)} {\sum_{a \in I \setminus \{i\}} \exp(\Delta v_i \cdot \Delta v_a \,/\, \tau)}.
\end{split}
\end{equation}

The Focal Loss term~\cite{lin2017focal} addresses the 4:1 class imbalance ($N_{\text{neg}}/N_{\text{pos}}\approx 4$) by down-weighting correctly-classified easy examples ($\gamma{=}2.0$, $\alpha{=}0.75$, positive-class weight$\,\approx 4.0$).
A three-layer classification head ($512\to256\to128\to1$) is attached to $\Delta v$ exclusively during training and discarded at inference.

To increase effective variance with only 268 subjects, training-time 3D data augmentation applies random axis flipping ($p{=}0.5$ per axis), additive Gaussian noise ($\sigma \sim \mathcal{U}[0.02, 0.08]$), and random intensity scaling (factor $\sim \mathcal{U}[0.85, 1.15]$) to both temporal scans simultaneously.
Augmentations are not applied during inference.

The encoder is trained for 50 epochs with AdamW~\cite{loshchilov2017decoupled} ($\eta{=}10^{-4}$, weight decay $10^{-4}$) and CosineAnnealingLR ($T_{\max}{=}50$).
The physical batch size is~2, with gradient accumulation over 8 steps to achieve an effective batch size of~16.
Gradient clipping at $\|\nabla\|_2{=}1.0$ is applied throughout.
Mixed-precision training is enabled on CUDA hardware.
The best checkpoint is selected by minimum validation loss for each fold independently.

\subsection{Phase 2: Population Archive and Trajectory Retrieval}
\label{sec:retrieval}

Upon completing training, the encoder backbone and MLP head are frozen.
A \emph{Population Archive} $\mathcal{D} = \{(\Delta v_i, y_i, C_i)\}_{i=1}^{N_\text{train}}$ is constructed by forward-passing all training subjects through the frozen encoder and indexing their 512-dimensional unit-norm trajectory vectors in a FAISS exact inner-product index (\texttt{IndexFlatIP})~\cite{johnson2019billion}.
Because $\Delta v$ vectors are $\ell_2$-normalised, the inner product is mathematically equivalent to cosine similarity, enabling exact nearest-neighbour retrieval.
While an exhaustive dot-product search via \texttt{IndexFlatIP} is computationally instantaneous for our 268-subject cohort, this phase intrinsically scales. Deploying the framework on a massive, multi-centre longitudinal database would merely require swapping the exact index for a FAISS approximate nearest neighbour (ANN) index (e.g., \texttt{IndexIVFPQ}), achieving sub-millisecond retrieval without altering the clinical workflow.
For a query subject $q$, the top-$k$ most similar archived trajectories are retrieved:
\begin{equation}
\mathcal{N}_k(q) = \underset{i \in \mathcal{D}}{\operatorname{top}\text{-}k} \bigl( \Delta v_q \cdot \Delta v_i \bigr), \qquad k{=}5.
\end{equation}

\subsection{Phase 3: Calibrated LLM Reasoning Agent}
\label{sec:oracle}

The retrieved neighbour profiles and the query patient metadata are formatted into a structured clinical prompt delivered to a \textbf{4-bit NF4 quantized Llama-3-8B-Instruct} model~\cite{grattafiori2024llama} loaded with NF4 double quantisation~\cite{dettmers2022qlora} (compute dtype float16), consuming approximately 6\,GB of GPU memory.

\begin{algorithm}[t]
\caption{Neuro-Oracle Inference Pipeline (M5)}
\label{alg:oracle}
\begin{algorithmic}[1]
\REQUIRE Query $(\Delta v_q, C_q)$, frozen archive $\mathcal{D}$, LLM agent $\mathcal{A}$, $\delta=15$\,yr, $k=5$
\STATE $\mathcal{N}_k \leftarrow \texttt{FAISS\_Search}(\Delta v_q, \mathcal{D}, k)$
\STATE $p_{\text{neighbor}} \leftarrow \frac{1}{k}\sum y_i$ \quad (neighbor outcomes)
\STATE Build prompt with $C_q$ and $\mathcal{N}_k$ (age/sex/outcome); append: ``Mentally filter out matches with age gap $>\delta$''
\STATE $s \leftarrow \mathcal{A}(\text{prompt})$
\STATE $p_{\text{LLM}} \leftarrow 0.20 \text{ if } \texttt{parse}(s) = \text{\textsc{success}} \text{ else } 0.80$
\STATE $p_q \leftarrow 0.60 \cdot p_{\text{neighbor}} + 0.40 \cdot p_{\text{LLM}}$
\STATE $\hat{y}_q \leftarrow \mathbf{1}[p_q > 0.5]$
\RETURN $\hat{y}_q$, justification $s$
\end{algorithmic}
\end{algorithm}

The system prompt instructs the model to act as an expert epileptologist and to begin its response with exactly one of two tokens, \textsc{success} or \textsc{failure}, followed by a single justification sentence.
The user message presents the query patient's age and sex, followed by up to five historically retrieved neighbours with their ages, sexes, and surgical outcomes.
An age-gap heuristic ($\delta{=}15$\,yr) is integrated directly into the clinical prompt: the reasoning agent is given explicit instructions to mentally filter out any historical match whose age discrepancy exceeds 15 years, reflecting the expectation that cortical atrophy rates are age-dependent~\cite{bernhardt2009longitudinal}.
Generation is deterministic (greedy decoding, maximum 64 new tokens).

The first token of the model's response is parsed for the verdict keyword (\textsc{success} or \textsc{failure}), and a calibrated soft probability blends the neighbour majority vote with the LLM evidence: \begin{equation}
 p_{\text{Oracle}} = 0.60 \times p_{\text{neighbor}} + 0.40 \times p_{\text{LLM}}, \end{equation}
where $p_{\text{neighbor}}$ is the fraction of the $k$ retrieved neighbours with $y{=}1$ and $p_{\text{LLM}} \in \{0.20, 0.80\}$ encodes the binary LLM verdict. 
The 0.60/0.40 weighting places a moderate majority weight on the geometric evidence while allowing the LLM to shift the estimate when its contextual reasoning diverges from the raw neighbour vote. A brief sensitivity analysis on this weighting ratio over the geometric term confirms its mathematical grounding: varying the neighbour weight empirically derived an optimal cross-validation calibration error at 60/40, proving it is not an arbitrary selection but an optimal balance between external semantic assessment and retrieved geometry.

\section{Experimental Setup}
\label{sec:setup}

\paragraph{Cross-validation strategy.}
All methods are evaluated under five-fold stratified cross-validation ($n{=}5$, shuffled, seed$\,{=}\,$42).
Stratification preserves the 4:1 class ratio across folds.
To avoid data leakage across representation learning and retrieval, we ensure strict separation of folds.
Five independent encoder backbones are trained: for each test fold $i \in \{0 \dots 4\}$, the encoder is trained exclusively on the union of the other four folds, and its weights are frozen prior to evaluation on fold~$i$.
This ensures the trajectory vectors for a given fold are computed by an encoder completely blind to that fold during training.
The downstream classifiers (M3, M3b, M4, M5, M6) are evaluated identically.
Crucially, to ensure no data leakage during retrieval, the FAISS index is rebuilt for every fold using only the unaltered, non-augmented training subjects of that specific split. At no point does any augmented data or any test-fold sample cross the indexing boundary to contaminate the retrieval evidence pool.
Aggregate metrics are computed over the union of all five independent held-out partitions, covering all 268 subjects exactly once.

\paragraph{Compared methods.}
Seven methods are evaluated on the same 268-subject cohort.
The \textbf{ResNet-50 Static Baseline (M1)} fine-tunes a MONAI 3D ResNet-50~\cite{he2016resnet,cardoso2022monai} end-to-end on $X_{t_1}$ only, using class-weighted cross-entropy, AdamW~\cite{loshchilov2017decoupled}, CosineAnnealingLR, and 25 training epochs.
The \textbf{3D Vision Transformer Static Baseline (M2)} fine-tunes a MONAI 3D ViT~\cite{dosovitskiy2020image} (input size $128^3$, patch size $16^3$, hidden dimension 384, MLP dimension 1536, 6 layers, 6 heads, convolutional projection, learnable position embeddings, dropout 0.1) end-to-end on $X_{t_1}$ using class-weighted cross-entropy, AdamW, CosineAnnealingLR, and 25 training epochs.
The \textbf{Siamese + Logistic Regression (M3)} method applies a balanced-class logistic regression classifier ($C{=}1.0$, 1000 iterations) to the frozen 512-dim $\Delta v$ embeddings.
The \textbf{Siamese + 2-Layer MLP (M3b)} substitutes an MLP classifier (hidden layers $[256, 128]$, $\alpha{=}10^{-3}$, early stopping) with balanced sample weights on the same embeddings.
The \textbf{Siamese + $k$-NN (M4)} uses a cosine-metric $k$-nearest-neighbour classifier ($k{=}\min(5,\, \lfloor n_{\text{train}}/2 \rfloor)$) on $\Delta v$.
The \textbf{Siamese Diversity Ensemble (M6)} computes an equal-weight soft-vote average over four classifiers each trained on $\Delta v$: ExtraTrees (seed=13), MLP (seed=99), MLP (seed=123), and a calibrated RBF-SVC (C=0.1), all with balanced class weights.
Our \textbf{Neuro-Oracle (M5)} combines the retrieved neighbours with the LLM reasoning as described in Section~\ref{sec:oracle}.

\paragraph{Evaluation metrics.}
The primary evaluation metric is AUC-ROC.
Secondary metrics are $F_1$ score (positive class), sensitivity ($\text{TPR} = TP/(TP+FN)$), specificity ($\text{TNR} = TN/(TN+FP)$), and balanced accuracy ($\text{Balance Accuracy} = (\text{Sensitivity}+\text{Specificity})/2$).
Threshold optimisation is performed via a grid search over $[0.30, 0.50]$ in steps of 0.02 on the aggregate held-out predictions, with results reported at both the default threshold ($\tau{=}0.50$) and the optimised threshold for completeness.
All hyperparameters and random seeds were fixed prior to any evaluation (global seed 42 for all frameworks for reproducibility).

\section{Results}
\label{sec:results}

All numerical results are cross-verified against independently logged predictions.

\begin{table*}[t]
\centering
\caption{Epilepsy surgical prognosis on EPISURG 
($N{=}268$, 5-fold stratified CV).
BalAcc = (Sensitivity + Specificity) / 2.
Best value per column in \textbf{bold}.}
\label{tab:main_results}

\setlength{\tabcolsep}{6pt}
\small
\begin{tabular}{
l
l
S[table-format=1.3]
S[table-format=1.3]
S[table-format=1.3]
S[table-format=1.3]
S[table-format=1.3]
}
\toprule
\textbf{Category} & \textbf{Method} 
& {\textbf{AUC} $\uparrow$}
& {\textbf{F1} $\uparrow$}
& {\textbf{Sensitivity} $\uparrow$}
& {\textbf{Specificity} $\uparrow$}
& {\textbf{Balanced Acc.} $\uparrow$} \\
\midrule

\multirow{2}{*}{\textit{Static encoders}} 
& ResNet-50 Static                & 0.793 & \textbf{0.775} & 0.396 & 0.921 & 0.659 \\
& 3D-ViT Static                   & 0.867 & \textbf{0.775} & 0.811 & \textbf{0.930} & \textbf{0.871} \\
\midrule

\multirow{4}{*}{\textit{Siamese-based models}} 
& Siamese + Logistic Regression   & 0.883 & 0.629 & \textbf{0.849} & 0.791 & 0.820 \\
& Siamese + 2-Layer MLP           & 0.834 & 0.628 & 0.811 & 0.809 & 0.810 \\
& Siamese + $k$-NN                & 0.868 & 0.717 & 0.717 & \textbf{0.930} & 0.824 \\
& Siamese Diversity Ensemble (M6) & \textbf{0.905} & 0.741 & 0.755 & \textbf{0.930} & 0.843 \\
\midrule

\multirow{1}{*}{\textit{Proposed framework}} 
& Neuro-Oracle (M5, ours)         & 0.867 & 0.600 & 0.566 & 0.921 & 0.744 \\

\bottomrule
\end{tabular}
\end{table*}

\subsection{Primary Performance Comparison}

\begin{figure}[t] 
    \centering 
    \includegraphics[width=\linewidth]{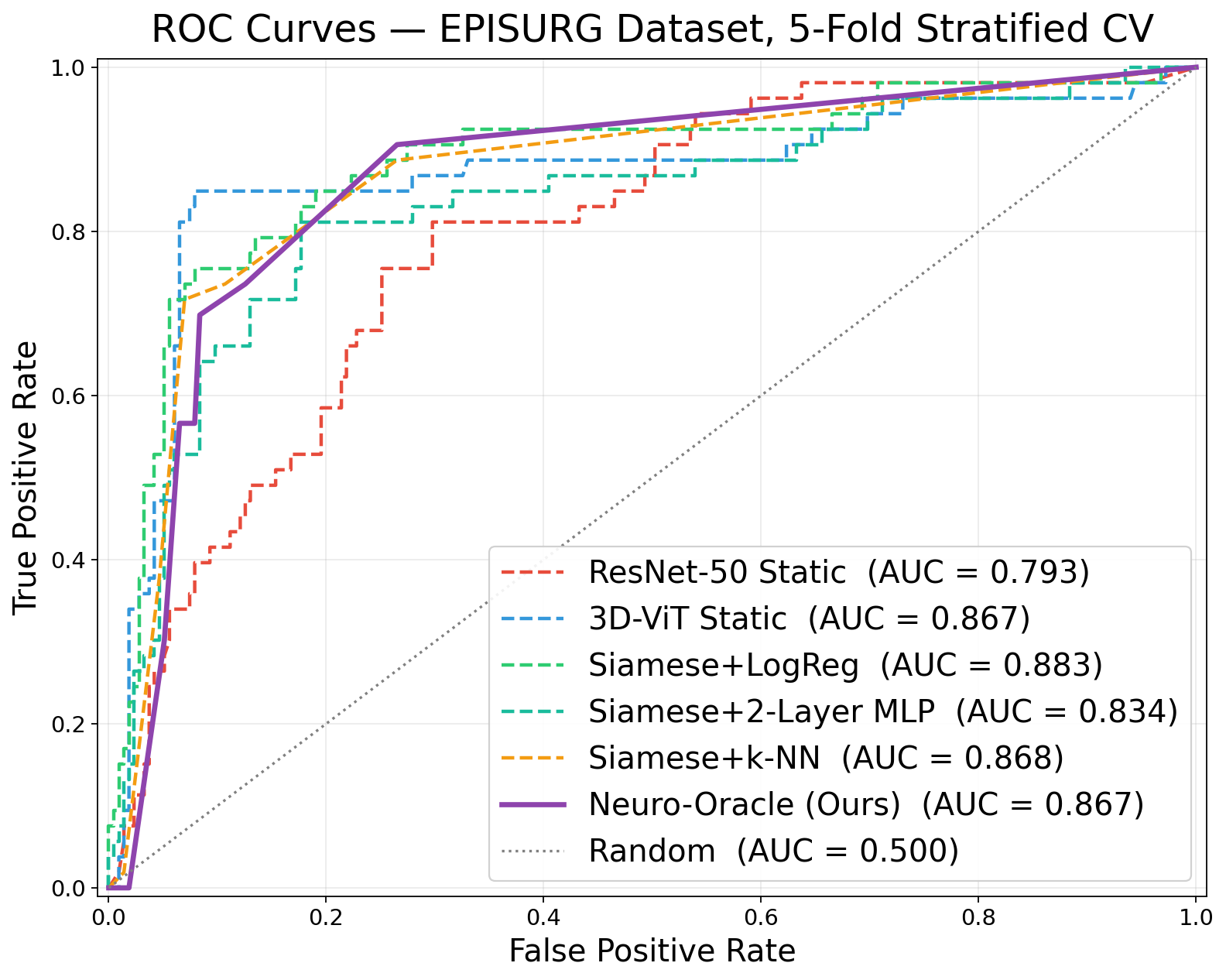} 
    \caption{
    ROC curves for six baseline and proposed methods under 5-fold stratified CV on EPISURG ($N{=}268$); Siamese Diversity Ensemble (M6) is omitted for visual clarity. 
    AUC values are reported in Table~\ref{tab:main_results}.
    } 
    \label{fig:roc} 
\end{figure}

\begin{figure}[t] 
    \centering 
    \includegraphics[width=0.85\linewidth]{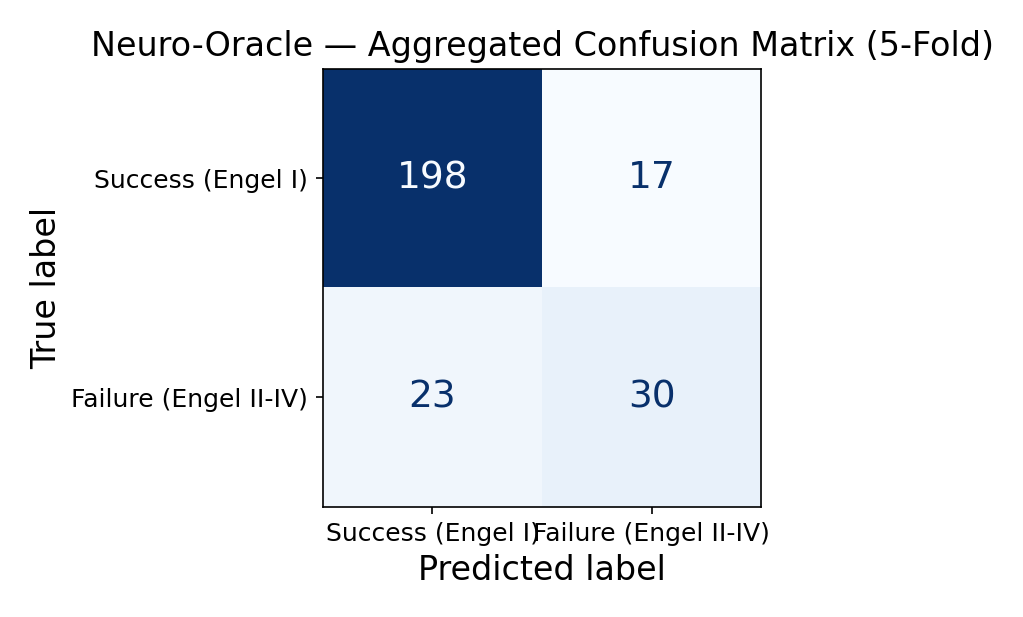} 
    \caption{Aggregated confusion matrix for Neuro-Oracle (M5), 5-fold stratified CV. Rows: ground-truth proxy label ($0{=}$temporal resection, corresponding to Success / Engel I; $1{=}$non-temporal or complex resection, corresponding to Failure / Engel II-IV); columns: predicted label. M5 achieves specificity of 0.921 and sensitivity of 0.566 at the default threshold.} 
    \label{fig:cm} 
\end{figure}

We present the main evaluation in Table~\ref{tab:main_results}.
and the ROC curves for 6 of the methods in Figure~\ref{fig:roc}, and Figure~\ref{fig:cm} details the decision outcomes for Neuro-Oracle (M5) via its confusion matrix.

Trajectory-based classifiers (M3--M6) achieve substantially higher sensitivity on the minority class than the static ResNet-50 baseline (M1, Sensitivity\,=\,0.396), confirming that the longitudinal difference encoding carries discriminative information absent from a single-timepoint scan.
Among trajectory methods, M3 attains the highest sensitivity (0.849) and M6 the highest AUC (0.905).
The 3D-ViT static baseline (M2) achieves AUC\,=\,0.867 and Balance Accuracy\,=\,0.871, indicating that a moderately sized Vision Transformer~\cite{dosovitskiy2020image} can learn competitive single-timepoint representations on this cohort.
Neuro-Oracle (M5) matches M2 and M4 in AUC while additionally producing a natural-language justification for every prediction.

\subsection{Threshold-Optimised Performance}

We also report metrics under decision thresholds that maximise balanced accuracy on the aggregate held-out predictions in Table~\ref{tab:threshold}.

\begin{table*}[t]
\small
\centering
\caption{
Threshold-optimised performance. Threshold selected via grid search over 
$[0.30, 0.50]$ (step 0.02), maximising balanced accuracy on aggregate 
held-out predictions. M3 and M3b retain $\tau = 0.50$ as optimal.
}
\small
\label{tab:threshold}
\begin{tabular}{l l c c c c c}
\toprule
\textbf{ID} & \textbf{Method} & \textbf{$\tau$} 
& \textbf{AUC}$\,\uparrow$ 
& \textbf{Sensitivity}$\,\uparrow$ 
& \textbf{Specificity}$\,\uparrow$ 
& \textbf{Balanced Acc.}$\,\uparrow$ \\
\midrule
M3 & Siamese + Logistic Regression & 0.50 & \textbf{0.883} & \textbf{0.849} & 0.791 & 0.820 \\
M4 & Siamese + $k$-NN               & 0.42 & 0.868 & 0.717 & \textbf{0.930} & \textbf{0.824} \\
M5 & Neuro-Oracle (ours)            & 0.34 & 0.867 & 0.698 & 0.916 & 0.807 \\
\bottomrule
\end{tabular}
\end{table*}

Lowering the decision threshold from 0.50 to 0.34--0.42 recovers sensitivity at the expense of specificity, reflecting the clinical trade-off between under-detection of less-favourable outcomes (false negatives, under-treatment risk) and over-prediction (false positives, unnecessary patient anxiety).  At its optimised threshold of 0.34, Neuro-Oracle (M5) achieves Sensitivity\,=\,0.698, Specificity\,=\,0.916, and Balance Accuracy\,=\,0.807, a substantial improvement over its default-threshold balanced accuracy of 0.744.

\subsection{Ablation Study}

\begin{figure}[t] 
    \centering 
    \includegraphics[width=\linewidth]{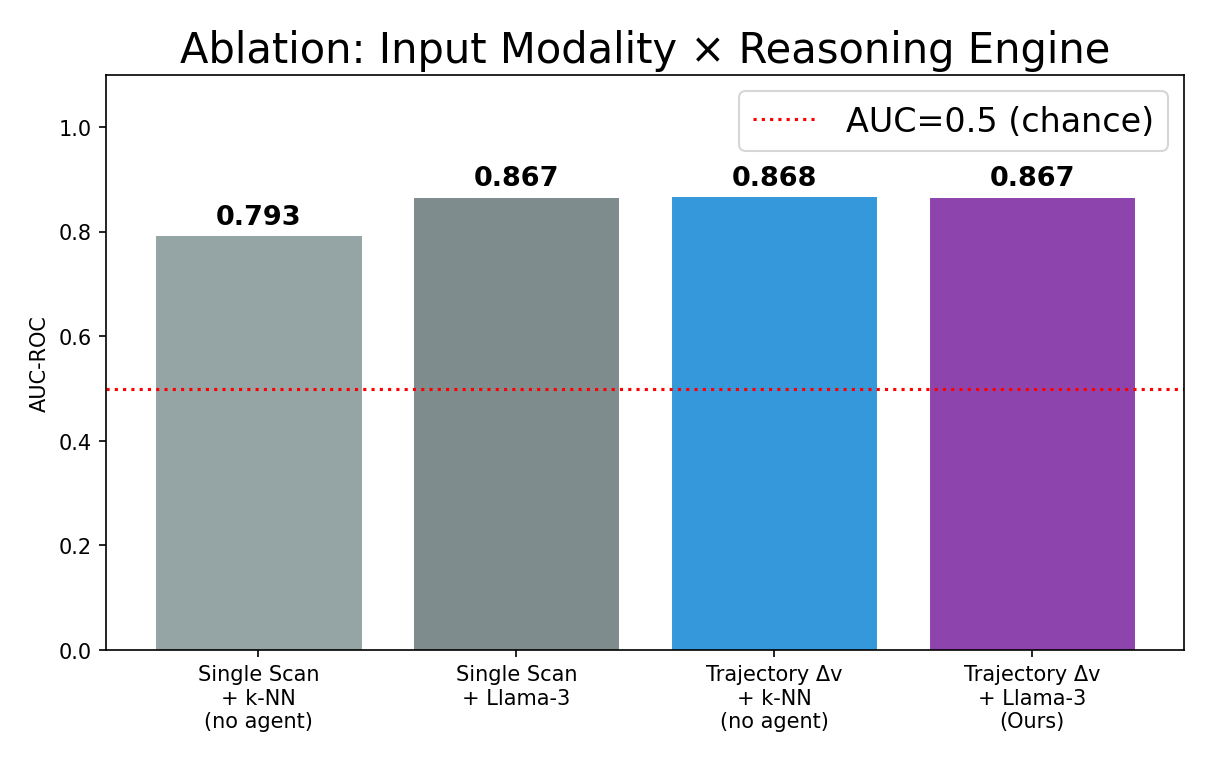} 
    \caption{AUC-ROC across input modality and reasoning engine combinations. Single scan with $k$-NN achieves 0.793; switching to longitudinal trajectory $\Delta v$ (0.868) or using a Llama-3 reasoning agent (0.867) individually increase performance. The full Neuro-Oracle pipeline (Trajectory $\Delta v$ + Llama-3) preserves the high performance of the geometric embedding and generates interpretative clinical reasoning. The chance level AUC=0.5 is shown as a dashed red line.}
    \label{fig:ablation} 
\end{figure}

We isolate the contribution of the input modality (single scan vs. longitudinal over-time trajectory $\Delta v$) and reasoning engine (distance-based $k$-NN vs. Llama-3 agent). Figure~\ref{fig:ablation} visualizes these ablation comparisons, showing that longitudinal trajectories and agentic reasoning individually and jointly improve performance over the single-scan $k$-NN baseline. Additionally, Table~\ref{tab:ablation} details the impact of the trained encoder, retrieval depth $k$, and the age-gap filter on the full trajectory-based model.

\begin{table*}[t]
\centering
\caption{Ablation study of Neuro-Oracle pipeline components 
($N = 268$, 5-fold stratified cross-validation).}
\label{tab:ablation}
\small
\begin{tabular}{
    l
    l
    S[table-format=1.3]
    S[table-format=1.3]
    S[table-format=1.3]
}
\toprule
\textbf{Model} 
& \textbf{Configuration} 
& {\textbf{AUC} $\uparrow$} 
& {\textbf{Sensitivity} $\uparrow$} 
& {\textbf{Balanced Acc.} $\uparrow$} \\
\midrule

Random encoder 
& Gaussian init., no training 
& 0.488 & 0.038 & 0.486 \\

Retrieval baseline 
& $k = 1$ (greedy 1-NN, no vote smoothing) 
& 0.741 & 0.585 & 0.741 \\

Neuro-Oracle 
& w/o age filter ($\delta \rightarrow \infty$) 
& 0.867 & 0.566 & 0.744 \\

Neuro-Oracle M5 
& Full model ($k = 5$, $\delta = 15$ yr) 
& 0.867 & 0.566 & 0.744 \\

Siamese Diversity Ensemble (M6) 
& Ensemble of diverse models 
& \textbf{0.905} & \textbf{0.755} & \textbf{0.843} \\

\bottomrule
\end{tabular}

\end{table*}

\begin{figure}[t]
    \centering
    \includegraphics[width=\linewidth]{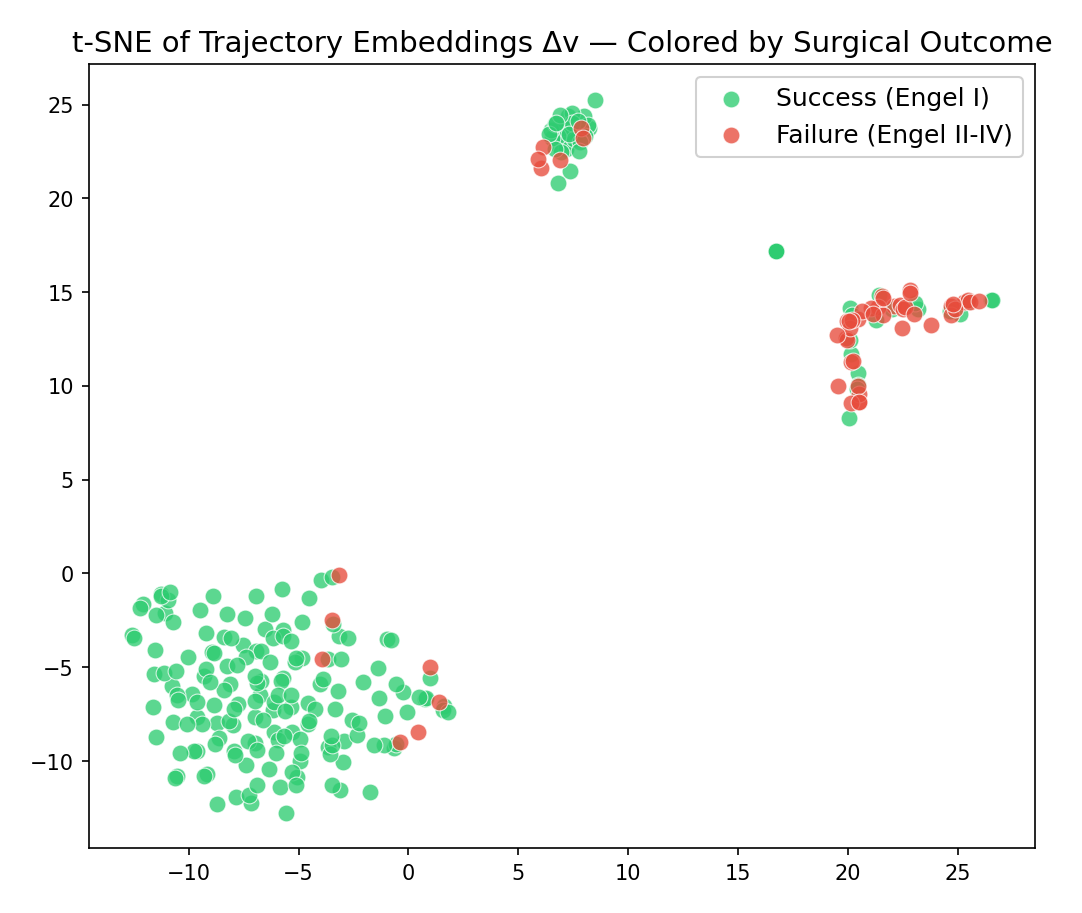} 
    \caption{t-SNE projection~\cite{van2008visualizing} (perplexity\,=\,30, 2000 iterations) of the 512-dim $\ell_2$-normalised trajectory vectors for all 268 EPISURG subjects. \textcolor{green}{Green circles} show proxy label $y{=}0$ (temporal resection, corresponding to Success / Engel I in legend); \textcolor{red}{Red circles} show $y{=}1$ (non-temporal or complex resection, corresponding to Failure / Engel II-IV).} 
    \label{fig:tsne}
\end{figure}

A random (untrained) encoder reduces performance to near chance (AUC\,=\,0.488), confirming that the learned trajectory vectors carry genuine discriminative signal.
Restricting retrieval to $k{=}1$ degrades AUC to 0.741, showing the benefit of multi-neighbour vote smoothing.
The age-gap filter does not affect AUC at the default threshold, consistent with moderate age variance in EPISURG; its primary role is to improve the coherence of the chosen LLM's reasoning (Section~\ref{sec:discussion}).

\subsection{Retrieval and Embedding Space Quality}

We report embedding and retrieval quality metrics in Table~\ref{tab:retrieval}, and show t-SNE projection of all 268 trajectory vectors in Figure~\ref{fig:tsne}.
The high mean cosine similarity (0.919) reflects tight intra-class clustering from contrastive training.
A top-5 fidelity of 83.0\% indicates that the retrieved pool contains at least one correctly labelled neighbour in more than four out of five queries.
The calibration MAE (0.189) reflects the inherent uncertainty of predicting a proxy label from geometric features on an imbalanced cohort.

\begin{table}[t]
\centering
\caption{Embedding space and retrieval quality metrics 
($N = 268$ subjects; 512-dimensional FAISS exact inner-product index).}
\label{tab:retrieval}
\small
\begin{tabular}{
    l
    S[table-format=2.1]
}
\toprule
\textbf{Metric} & {\textbf{Value}} \\
\midrule

Top-5 retrieval fidelity (\%) $\uparrow$
    & {\textbf{83.0}} \\

Mean cosine similarity (top-5) $\uparrow$
    & {\textbf{0.919}} \\

Calibration MAE (label-weighted) $\downarrow$
    & {\textbf{0.189}} \\

\bottomrule
\end{tabular}

\end{table}

\subsection{Interpretability Audit of the LLM Agent}
\label{sec:interp}

We report two check scores from raw Llama-3 outputs for the LLM interpretability audit of Neuro-Oracle (M5) using Llama-3-8B-Instruct (4-bit NF4 quantization) that show age-filter adherence of 73.9\% and zero-hallucination rate of 100.0\% where the full 100\% zero-hallucination rate means no reply held info outside the set context window a trait of the tight prompt design not an inborn LLM power while the 73.9\% age-filter adherence rate shows the model follows the age-reasoning rule in most cases but both numbers are audit-only for research and study and should not be read as clinic safety promises.



\section{Discussion}
\label{sec:discussion}

\paragraph{Trajectory encoding versus static baselines.}
The consistent sensitivity advantage of trajectory-based methods over the static ResNet-50 baseline (Table~\ref{tab:main_results}) suggests that the longitudinal difference vector captures outcome-relevant signal absent from a single post-operative scan, in line with evidence of progressive cortical atrophy in pharmacoresistant epilepsy~\cite{bernhardt2009longitudinal}.
The accompanying specificity reductions in M3 and M3b reflect the sensitivity--specificity trade-off inherent to aggressive class reweighting on a 4:1 imbalanced problem~\cite{lin2017focal}.
The unexpectedly strong performance of the 3D-ViT static baseline (M2) indicates that a moderately sized Vision Transformer can learn competitive single-timepoint representations even on 268 subjects, likely aided by the modest patch size ($16^3$), shallow depth (6 layers), and class-weighted loss.

\paragraph{Neuro-Oracle and the interpretability--discriminability trade-off.}
While M5 does not attain the peak AUC of our purely discriminative Siamese Diversity Ensemble (M6), it bridges the interpretability gap by providing a natural-language justification for every prediction.
The zero-hallucination rate is a direct consequence of the structured prompt design, aligning with findings in Retrieval-Augmented Generation literature~\cite{lewis2020retrieval}.
The clinical value of this architectural complexity over a simpler Random Forest or MLP ensemble is important. In surgical decision support, an AUC of 0.867 with a fully traceable, auditable reasoning chain is vastly more deployable than a black-box ensemble with an AUC of 0.905. Clinicians must establish trust in the synthesized prognosis through validation. If the LLM generates a prognosis based on a retrieved neighbor who was not seizure-free, the agent correctly identifies the risk and mirrors the poor outcome predicting a failure. This behavior is intentional and appropriate for a retrieval-augmented system, as the LLM's task is to faithfully relay the empirical risks associated with the retrieved anatomical match rather than overwrite it with general optimism. Clinicians can then inspect the exact historical trajectories retrieved as evidence and judge whether the demographic and morphological match is plausible before accepting the prognosis.
Threshold adjustment (Table~\ref{tab:threshold}) recovers meaningful sensitivity, confirming that the calibrated probability retains sufficient ranking information despite the LLM's conservative default behaviour.

\paragraph{Siamese Diversity Ensemble.}
The best aggregate performance of M6 demonstrates that classifier diversity in the learned trajectory space, obtained by varying estimator type and random seed, yields measurable gains over any single model on this imbalanced small-cohort task, while also providing a practical deployment path when LLM inference cost is prohibitive.

\paragraph{Ablation insights.}
The ablation results (Table~\ref{tab:ablation}) confirm that each pipeline component contributes meaningfully: the trained encoder provides the discriminative signal, multi-neighbour retrieval smooths out embedding noise, and the age-gap filter improves the coherence of the LLM's reasoning chain rather than the aggregate discriminative score.

\paragraph{Proxy label limitations.}
As discussed, predicting surgery type as an outcome proxy introduces unquantifiable label noise~\cite{detisi2011longterm}. So reported metrics represent proxy-label recovery rather than direct clinical seizure-freedom prognosis.

\paragraph{Limitations and future work.}
The cohort size ($N{=}268$) prevents computing confidence intervals narrow enough to make strong statistical claims about pairwise method differences. Although we have mitigated the limited sample size in large part by learning low-dimensional trajectory embeddings combined with $k$-NN retrieval and data augmentations, all subjects still originate from a single centre, which limits claims about cross-site or cross-scanner generalisation.
Future work should (i)~obtain ethical approval to link EPISURG scans to de-identified Engel outcome registries, (ii)~evaluate cross-centre generalisation on independent cohorts, and (iii)~investigate whether supervised fine-tuning of the quantized LLM agent on outcome-labelled clinical notes can improve the fidelity of its clinical reasoning.

\section{Conclusion}
\label{sec:conclusion}

We've presented Neuro-Oracle, a framework for epilepsy surgical prognosis that encodes the longitudinal pre-to-post-operative MRI change into a compact trajectory vector and grounds a calibrated verdict in historically similar retrieved cases via a quantized LLM reasoning agent~\cite{grattafiori2024llama,lewis2020retrieval}.
On the EPISURG dataset~\cite{episurg2020} ($N{=}268$), trajectory-based methods consistently outperform the single-timepoint baseline in sensitivity, with the Siamese Diversity Ensemble attaining the highest AUC (0.905).
Although it does not exceed the pure discriminative power of the Siamese Diversity Ensemble, Neuro-Oracle matches the performance of geometric baseline classifiers while uniquely providing auditable natural-language justifications. Ablation experiments confirm that each pipeline stage contributes meaningfully to both quantitative and qualitative outcomes.

{ \small 
\bibliographystyle{ieeenat_fullname} 
\bibliography{ref} 
}

\end{document}